# Effects of Dark Matter in Red Giants

Clea Sunny[1] • Arun Kenath*[1,2] • C Sivaram[3] • S B Gudennavar[1]

**Abstract**: Dark matter (DM) which constitutes five-sixths of all matter is hypothesised to be a weakly interacting non-baryonic particle, created in the early stages of cosmic evolution. It can affect various cosmic structures in the Universe via gravitational interactions. The effect of DM in main sequence stars and stellar remnants like neutron stars and white dwarfs has already been studied. Red giant phase is a late stage of the evolution of stars. In this work, we study, low-mass red giants stars with admixture of DM and how this can effectively change the intrinsic properties of red giants such as their luminosities, temperatures and lifetimes.

---

*Corresponding author

[1]Department of Physics and Electronics, Christ (Deemed to be University), Bengaluru, 560029, Karnataka, India

[2]Department of Physics, Christ Junior College, Bengaluru, 560029, Karnataka, India

[3]Indian Institute of Astrophysics, Bengaluru, 560034, Karnataka, India

E-mail addresses: clea.sunny@phy.christuniversity.in (Clea Sunny); kenath.arun@cjc.christcollege.edu (Arun Kenath); sivaram@iiap.res.in (C Sivaram); shivappa.b.gudennavar@christuniversity.in (S B Gudennavar)

# 1 Introduction

The measurements from Planck data [1] provide information on the total mass content of the Universe. DM is about five times more abundant than normal baryonic matter. It is a non-luminous matter (neither absorbing nor emitting radiation), which only interacts gravitationally and thus binds galaxies together. For the past few decades, the idea that Universe consists solely of baryonic matter has been challenged due to more evidence that suggest the existence of DM. Galactic flat rotation curves, gravitational lensing, fluctuations in cosmic microwave background (CMB), velocity dispersions, X-ray measurements from galaxy clusters, etc. are some of the indirect observational evidence which show the presence of DM [2].

Although it makes 26.8% of total energy density of the Universe, its nature still remains a mystery. By making accurate measurements of CMB fluctuations, we can determine the density of the universe. All these observations concerning the existence of DM lead us to plethora of ideas regarding its influence on various astronomical objects including stars in their late stages of evolution. The effect of DM on main sequence stars and stellar remnants such as neutron stars and white dwarfs has already been studied [3-5].

The very recent observations on the role of dark matter in main sequence stars have shown that at the earlier epoch, the Universe would have had higher density of DM particles which would have contributed to star formation in the early Universe [3]. The presence of DM admixture with baryonic particles in the star increases its core mass and thereby increases its temperature and luminosity. As the core mass increases, thermonuclear reactions occur in a faster rate which lead to decrease in lifetime of the star. This effect of DM particles on main sequence stars has been extended to evolved stars as well. The discrepancy in the observed and theoretical mass limit of neutron stars (NS) was addressed using this model (admixture of DM) [4].

Observations have indicated that we do not see neutron stars of mass near the theoretical upper limit as predicted. One of the possible explanations in the discrepancy could be the presence of DM particles. A small admixture of DM particles at the core lowers the theoretical upper mass limit. Similar to the case of neutron stars, the mass limit for the white dwarfs (WD) varies inversely with the square of the constituent particle mass [5]. If the constituents are heavier (as in the case with admixture of DM particles), the upper mass limit of white dwarfs would be lowered. Since DM particle mass is much greater than that of baryonic particle mass, presence of even 1% of DM particles can lower the upper mass limit of white dwarfs. This model was used to account for sub-luminous Type Ia supernovae without invoking dark energy.

In this work, we extend this idea and study how DM admixture can affect the stars in its red giant phase, in particular, the case of low-mass red giant stars. We study the variations in the properties of these stars due to the admixture of DM and how it can also lead to some insights into the nature of DM.

**2 Red giants and DM**

The end of the main sequence phase of evolution occurs when hydrogen burning ceases in the core of a star. After the hydrogen fuel in the core is nearly exhausted, the core becomes isothermal (temperature gradient, $\frac{dT}{dr} = 0$) and thus the nuclear luminosity vanishes at the core since, $L(r) \propto \frac{dT}{dr}$. The core without a temperature gradient cannot provide the pressure gradient to support the matter against the gravity and hence the core contracts [6]. Helium is building up in the core and as it does the density and temperature of the core increases. This heat is not sufficient for helium fusion to occur inside the core but it is enough for hydrogen fusion in a shell surrounding the core. This will increase the temperature of the shell and thus it leads to the expansion and cooling down of the outer layers of the star. This is the red giant phase of a star. With the envelope expansion and decrease in temperature, the photospheric opacity increases. The result is that a convection zone is developed near the surface for both low- and intermediate-mass stars [6].

The evolutionary track of stars is based on their masses. The evolution of low- (typically $M < 2.3 M_\odot$) and intermediate-mass stars ($M > 2.5 M_\odot$) is quite different from that of massive stars ($M > 8 M_\odot$). This is because low-mass stars generate energy from p-p chain and have no convective cores, unlike massive stars. Hence the material is not mixed within the core of the low-mass stars [7].

Low-mass stars have central temperatures and densities closer to the degeneracy limit than high-mass stars. Because the cores are nearly degenerate, the Chandrasekhar-Shoenberg mass limit is fairly irrelevant for low-mass stars [7]. For massive stars, the entire material is mixed up and it is very difficult to obtain relations for various properties of the star. Various authors have published core mass-luminosity relations for AGB stars based on their complete solutions of the stellar structure equations using homology approximations [8].

Studies [2] have been done on the influence of dark matter particles on various cosmic structures. If we observe closely the variations in the properties of these structures occurring due to unknown reasons, there is a slight possibility that it might reveal the presence of DM in them. We now look at the scenario where the presence of DM in red giant stars can cause detectable features on its intrinsic characteristics. We also look at observational signatures in such objects.

**3 Effect of DM on luminosities of low-mass red giants**

The core contraction does lead to the expansion of the hydrogen-rich envelope outside the shell source and to a significant increase in the luminosity. This part of the evolution is essentially controlled by the core mass $M_c$ and is independent of the mass of the envelope $(M - M_c)$. Therefore the luminosity of a hydrogen burning shell surrounding a degenerate helium core of mass $M_c$ is roughly a function of core mass only.

In 1970, Refsdal and Weigert developed homology-type relations which are very helpful in understanding the properties of shells and core of a star. Kippenhahn [9] extended their work and has provided relations for low-mass stars. In particular, they all show analytically that, as long as radiation-pressure can be neglected, there exists a core-mass luminosity relation [7] of the form,

$$L \propto M_c^7 \qquad (1)$$

For red giants, the outer envelope is formed due to thermal pressure. Hence the role of gravity is negligible for the envelope, since DM only interacts with gravity. Thus the effect of DM is more in the core of red giants. For low-mass red giant stars, we can use the core mass-luminosity relation given in equation (1) to check the effect of DM fraction in the luminosity. The total mass can be written in terms of DM as,

$$M_T = M_B + M_{DM} = (1 + f)M_B \qquad (2)$$

where $M_{DM} = fM_B$, i.e. DM particles present is assumed to be a fraction of baryonic particles.

For low-mass stars, $M_B = M_c$ the core mass. Therefore we can write the total mass as,

$$M_T = M_c + M_{DM} \qquad (3)$$

Using equation (1) we can derive luminosities due to the presence of DM ($L_{DM}$) and normal baryonic matter ($L_0$). When we take the ratio of luminosities, we get

$$\frac{L_{DM}}{L_0} = (1 + f)^7 \qquad (4)$$

For the fraction of DM (given by the parameter $f$), the change in ratio of luminosity is observed as follows:

**Table 1** Change in ratio of luminosity with admixtures of DM

| $f$ (%) | $L_{DM}/L_0$ |
|---|---|
| 10 | 1.949 |
| 20 | 3.583 |
| 30 | 6.275 |
| 40 | 10.541 |
| 50 | 17.086 |
| 60 | 24.843 |
| 70 | 41.034 |
| 80 | 61.222 |
| 90 | 89.387 |
| 100 | 128 |

The variation in luminosity of red giants due to DM to that in the absence of DM ($L_0$) is plotted in figure 1. As expected, more the fraction of DM particles in the core of red giants, more luminous the star becomes.

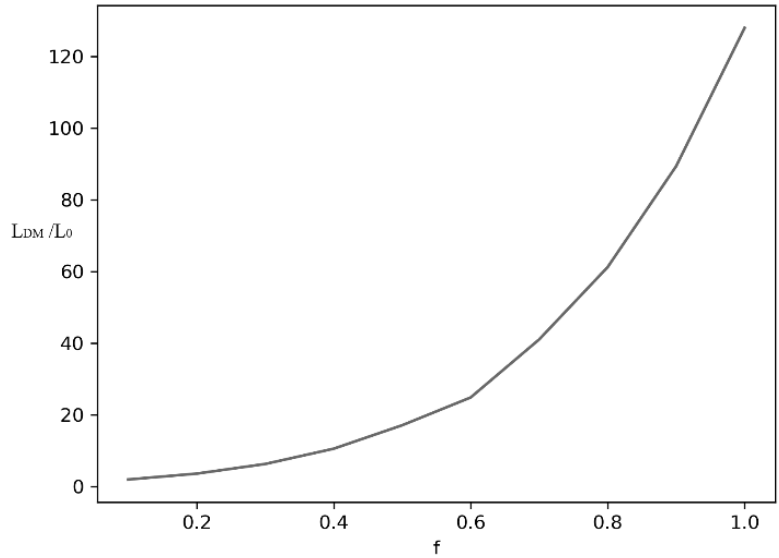

**Figure 1** Change in luminosity with increasing DM fraction

## 4 Effect of DM on temperatures of low-mass red giants

As the helium core begins to contract, the radius of the star increases slowly. The contraction increases the temperature sufficiently to ignite hydrogen in a shell. As the energy production switches from core to shell burning, the core mass starts increasing with the ashes of shell burning added to it. With growing core mass, the temperature of the core rises. When the core mass is $M_c \approx 0.45 M_\odot$, the core temperature reaches $T_c \approx 10^8 K$, at which the helium is ignited. The ignition of helium burning occurs initially in a shell around the centre, but the core quickly becomes involved. This process is called the helium flash. Stellar evolution calculations of low-mass stars are often terminated at this point due to the very rapid rate of helium core flash. When the nuclear reactions are triggered in fully degenerate matter, the increase in temperature will lead to a further increase in the rate of energy production, quickly leading to a runaway situation. The temperature of the region goes up rapidly until the electrons become non-degenerate.

The equation of state outside the core can be taken to be that of an ideal gas. Under such conditions, homologous solutions will exist for the equations of stellar structure in which the core mass is taken to dominate over the envelope mass; i.e. we can replace $M(r)$ with $M_c$ in the equation of hydrostatic equilibrium. Straightforward analysis now leads to the relations [7];

$$T \propto \frac{M_c}{R_c} \tag{5}$$

But the core has a factor $f$ of DM in it. Hence we have the temperature in terms of DM fraction as:

$$T_{DM} \propto (1+f)\frac{M_c}{R_c} \tag{6}$$

The ratio of the temperature in the presence of DM particles $(T_{DM})$ to that in the absence of DM particles $(T_0)$ is:

$$\frac{T_{DM}}{T_0} = (1+f) \tag{7}$$

Table 2 Change in ratio of temperatures with admixture of DM

| $f$ (%) | $T_{DM}/T_0$ |
|---|---|
| 10 | 1.1 |
| 20 | 1.2 |
| 30 | 1.3 |
| 40 | 1.4 |
| 50 | 1.5 |
| 60 | 1.6 |
| 70 | 1.7 |
| 80 | 1.8 |
| 90 | 1.9 |
| 100 | 2 |

Thus a DM fraction of $f \sim 0.1$ will lead to a temperature increase of 1.1. The ratio of core temperature with increasing DM fraction to $T_0$ is plotted in figure 2.

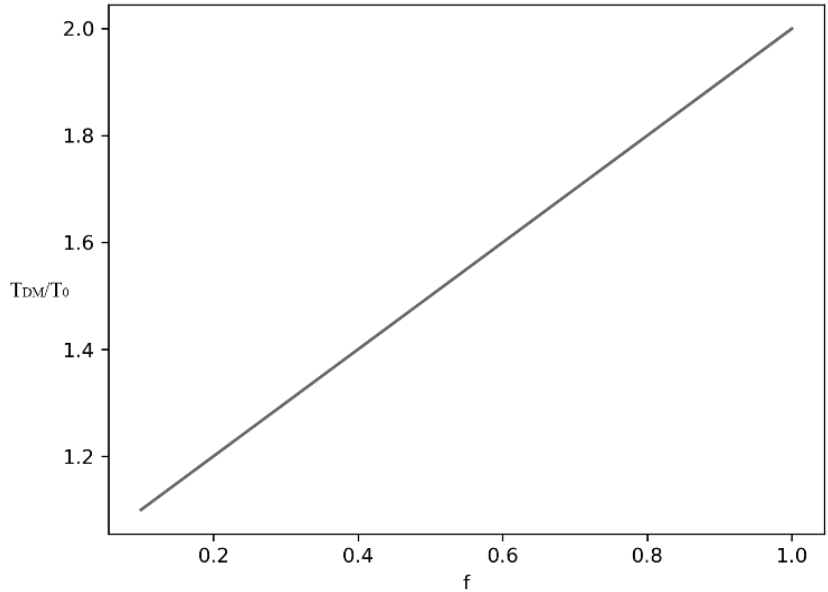

**Figure 2** Change in temperature with increasing DM fraction

The energy generation in low-mass stars having lower temperatures is dominated by p-p chain mechanism during their hydrogen burning evolution, whereas more massive stars with their higher central temperatures convert hydrogen to helium by CNO cycle. The nuclear reactions involving p-p chain mechanism has a very weak temperature dependence. When the core mass exceeds Chandrasekhar-Schonberg mass limit, the core collapses on the Kelvin-Helmholtz time-scale [6]. Initially for low-mass stars ($M <$ around $1.8 M_\odot$), the core is not hot enough to ignite helium inside the core.

But as the helium core continuous to collapse and become completely degenerate, the temperature ($10^8 K$) and density become high enough to initiate the triple-alpha (3α) process and the resulting energy release is almost explosive which only lasts for a few seconds. The origin of the explosive energy release is in the very weak temperature dependence of electron degeneracy pressure and the strong temperature dependence of triple-alpha process [6]. The energy generation rate for the triple-alpha process is given by,

$$\epsilon_{3\alpha} \propto \rho^2 Y^3 f_{3\alpha} T_8^{41} \tag{8}$$

where $\rho$ is density, $Y$ is mass fraction of helium, $f_{3\alpha}$ is the screening factor for triple-alpha process and $T_8$ is a dimensionless expression of temperature in units of $10^8 K$ [6].

This strong temperature dependence of energy generation during helium burning in low-mass red giant stars with the DM-temperature relation given in equation (7) can be used together to check how DM can cause an additional variation in temperature while helium burns in the core. From equation (8), we have the energy reaction rate:

$$(\epsilon_{3\alpha})_0 \propto T^{41} \tag{9}$$

If there are DM particles present in the core of the star, it would affect the temperature of the core as we have already calculated in equation (7). Using this equation, the energy generation rate in the presence of DM has the form:

$$(\epsilon_{3\alpha})_{DM} \propto (1+f)^{41} T^{41} \tag{10}$$

Again we have the ratio of the reaction rates in the presence of DM, $(\epsilon_{3\alpha})_{DM}$, to that in the absence of DM, $(\epsilon_{3\alpha})_0$, given as:

$$\frac{(\epsilon_{3\alpha})_{DM}}{(\epsilon_{3\alpha})_0} = (1+f)^{41} \tag{11}$$

Here we can see that a DM fraction of $f \sim 0.1$ will lead to an increase in energy generation rate of 49.79. This implies that since temperature is higher in the presence of DM, the reaction rates will be higher and hence helium will burn faster in much shorter time.

## 5 Effects of DM on the lifetimes of red giants

The conversion of four hydrogen nuclei to one helium nucleus releases $\sim$26.7 MeV, of which a small fraction is taken away by the neutrinos, leaving $\sim$25 MeV trapped in the star. If a fraction $m$ of the total mass of the star can be converted during the main-sequence phase, then the main-sequence lifetime will be, $\tau_{ms} \approx \eta m \left(\frac{Mc^2}{L_{max}}\right)$, where thermonuclear efficiency $\eta = 0.007$ for main sequence stars. In the initial phase $\sim$10% of stellar mass gets converted, leading to $\tau_{ms} \approx 10 \, Gyr (M/M_\odot)(L/L_\odot)^{-1}$. All other phases of evolution last for much shorter period of time.

For example, stars with an initial mass less than approximately $1.8 - 2.2 \, M_\odot$ experience helium flash with a luminosity of $\sim 50 \, L_\odot$. The helium core of these stars have masses of $\sim 0.45 \, M_\odot$. Approximately half of this helium is converted into carbon and the other half becomes oxygen, releasing $7.2 \times 10^{-4} (\Delta M) c^2$ joules of energy. Hence the time scale is approximately constant and is given by

$$\tau_{rg} \approx 7.2 \times 10^{-4} \frac{0.45 M_\odot c^2}{50 L_\odot} \approx 0.5 \, Gyr, \tag{12}$$

which is quite short compared with the main-sequence lifetime [10].

In the case of low-mass stars, analysing homologous solutions for the equations of stellar structure will provide a relation for luminosity which depends on core mass and core radius, given as:

$$L \propto M_c^7 R_c^{-16/3} \tag{13}$$

The above relation can be used to determine the evolutionary track of the star in HR diagram [7]. We know that, for the low-mass red giant, the mass-luminosity relation is given by equation (1). Therefore, the maximal luminosity of red giants has a factor of $(1+f)^7$, with the presence of DM. Since the luminosity of such a star is increased by the DM particles, their lifetime will be correspondingly reduced with an increasing admixture of DM particles. The lifetime of red giants in the absence of DM is given by,

$$\tau_0 = \frac{\eta M_c c^2}{L_{max}} \tag{14}$$

where $\eta = 7.2 \times 10^{-4}$ is the thermonuclear efficiency of the nuclear reactions in the red giants.

So now we can write the lifetime of red giants in the presence of DM as,

$$\tau_{DM} = \frac{\eta (1+f) M_c c^2}{M_T^7 R_c^{-16/3}} \tag{15}$$

$$\Rightarrow \frac{\tau_{DM}}{\tau_0} = (1+f)^{-6} \tag{16}$$

i.e. ~0.1 of DM reduces the life time of red giant phase of such stars by a factor of 0.564. For each percentage increase of DM by a factor $f$, the relative change in lifetime is given as follows.

**Table 3** Change in ratio of lifetimes with admixtures of DM

| $f$ (%) | $\tau_{DM}/\tau_0$ |
|---------|---------------------|
| 10      | 0.564               |
| 20      | 0.335               |
| 30      | 0.207               |
| 40      | 0.133               |
| 50      | 0.0881              |
| 60      | 0.0596              |
| 70 | 0.0414 |
| 80 | 0.0294 |
| 90 | 0.0213 |
| 100 | 0.0156 |

The variation of lifetime with increasing DM fraction to the lifetimes without any DM admixture $\tau_0$ is plotted in figure 3.

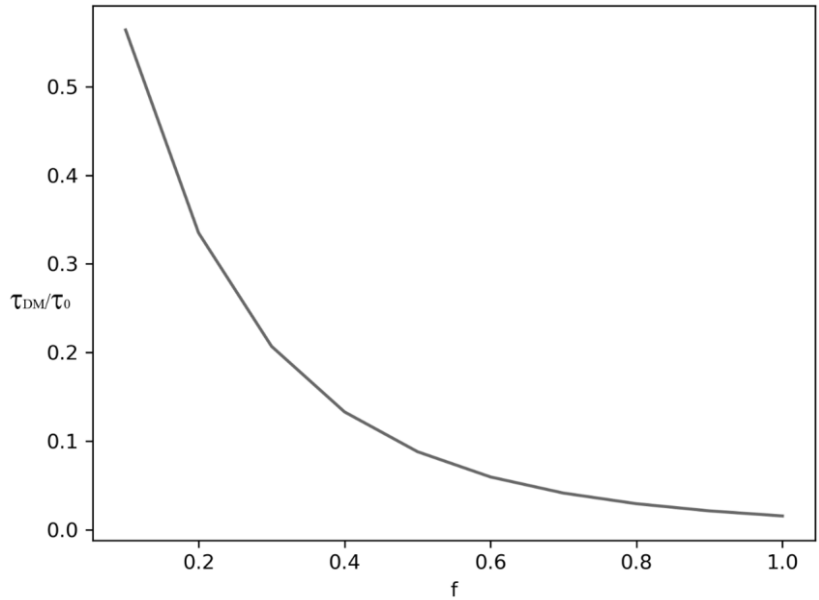

**Figure 3** Change in lifetime with increasing DM fraction

## 6 Summary and Conclusions

Evolved stars show many variations in their physical parameters over the time scale when they move out of the main sequence branch. Apart from the usual changes, variations in physical processes will be much more prominent if dark matter exists inside stars and in evolved stars. Red giant phase is a late stage of stellar evolution for low- and intermediate-mass stars where temperature gradient inside a star causes visible changes in the overall appearance of the star. Here we considered the case of only low-mass red giants.

From the calculations and graphs discussed above for a red giant, we can conclude that in the case of low-mass red giants, as core mass increases in the presence of dark matter, the luminosity increases by a large factor and also the temperature of the core rises linearly with the mass. Since the luminosity is higher, the lifetime of the red giant star reduces to nearly half in the presence of dark matter. This implies that, the star will remain in the red giant phase only for an even smaller fraction of its lifetime. These calculated lifetimes in the red giant phase can cause observational conflicts with the currently obtained lifetimes from the usual stellar evolution studies. These could provide stringent constraints on the presence of DM in the interiors of these stars. For instance, a 10 percent fraction of DM can cause the lifetime to be reduced by a factor of 2. Another evidence for the existence of DM would be the strong temperature dependence of helium burning during helium core flash. Thus from what we have discussed above, we conclude by pointing out that these variations can be the observational signatures for possible presence of dark matter particles inside stars.

Finally, it is worth remarking that the work presented in this paper could be usefully employed in the evolution models of high-mass red giant stars, which fuses heavier elements such as oxygen, nitrogen and iron in its core, with the admixture of DM in them. In the future, it may provide more exciting promises.